\documentclass[showpacs,showkeys,amsmath,amssymb]{revtex4}
\usepackage{rotating}
\usepackage{graphicx}
\usepackage{bm}
\begin{document}

\title{Deformed shell model results for two neutrino positron double beta
decay of $^{74}$Se}

\author{A. Shukla$^{1,2}$, R. Sahu$^3$,  and V.K.B. Kota$^{1,4}$}

\affiliation{$^1$Physical Research Laboratory, Ahmedabad 380 009,  
India \\ $^2$Rajiv Gandhi Institute of Petroleum Technology, Raebareli 
229 316, U.P., India \\ $^3$Physics Department, Berhampur University, 
Berhampur 760 007, Orissa, India \\ $^4$Department of Physics, Laurentian 
University, Sudbury, ON P3E 2C6, Canada} 

\begin{abstract}

Half-lives $T_{1/2}^{2\nu}$ for two neutrino positron double beta decay
modes $\beta^{+}$EC/ECEC are calculated for $^{74}$Se, a nucleus of current
experimental interest, using  deformed shell model based on Hartree-Fock
states employing a modified Kuo interaction in ($^{2}p_{3/2}$,
$^{1}f_{5/2}$, $^{2}p_{1/2}$, $^{1}g_{9/2}$) space. The calculated
half-life for the ECEC mode is $\sim 10^{26}$yr and it may be possible to
observe this in the near future with improved sensitivity of experiments.

\end{abstract}

\pacs{23.40.-s, 23.40.Hc, 21.60.Jz, 27.50.+e}

\keywords{positron double beta decay, Deformed shell model, two neutrino
$\beta^+\beta^+/\beta^+$EC/ECEC decay modes, half lives, $^{74}$Se}

\maketitle

\section{Introduction}

Double beta decay (DBD) is a rare weak interaction process in which two
identical nucleons inside the nucleus undergo decay with or without
emission of neutrinos. The two neutrino double beta decay ($2\nu \beta
^{-}\beta ^{-}) $ which was first predicted long back by Meyer \cite{y1} is
fully consistent with standard model and has been observed experimentally
in more than 10 nuclei. The neutrinoless double beta decay ($0\nu \beta
^{-}\beta ^{-})$ which involves emission of two electrons and no neutrinos,
has not been observed experimentally and it violates lepton number 
conservation. The claim for observation of $0\nu \beta^{-}\beta ^{-}$
decay  of  $^{76}$Ge by Heidelberg-Moscow group \cite{kla01} is
controversial and yet to be confirmed by other ongoing experiments. This
process is one of the best probes for studying physics beyond the standard
model. To extract the mass of neutrino via $0\nu \beta^{-}\beta ^{-}$ 
decay,  it is necessary to have good nuclear structure models for
calculating the nuclear transition matrix elements (NTME) involved. Large
number of theoretical studies for various candidate nuclei, for $2\nu$ and
$0\nu \beta^{-}\beta^{-}$ decay, using many nuclear models have been
carried out  so that the calculated NTME can be established to be reliable;
see for example \cite{y2, y3}. It is important to add that all models will
not work for all DBD nuclei due to various reasons. For example for $A >>
64$ nuclei full shell model in ($^{2}p_{3/2}$, $^{1}f_{5/2}$,
$^{2}p_{1/2}$,  $^{1}g_{9/2}$) space is still not feasible.

In contrast to the 2$\nu\; \beta ^{-}\beta ^{-}$ decay, the positron decay
modes, i.e. $2\nu\; \beta ^{+}\beta^{+}/\beta^{+}$EC/ECEC decay modes 
(hereafter, all these three combined is called $2\nu\;e^+$DBD) are not yet
observed experimentally and hence there are not many theoretical studies of
the NTME involved in $2\nu\;e^+$DBD. However, in the last few years serious
attempts are made to measure half-lives for $2\nu\;e^+$DBD modes in the
upper $(pfg_{9/2})$ shell nuclei $^{78}$Kr \cite{kr78},  $^{64}$Zn
\cite{zn64} and $^{74}$Se \cite{se74} (in the past, attempts are  also made
for $^{106,108}$Cd \cite{cd106} and $^{130,132}$Ba \cite{ba130} nuclei). 
Prompted by this experimental interest, recently \cite{y5} we have carried
out calculations for $^{78}$Kr using the so called deformed shell model
(DSM) by employing a modified Kuo interaction in ($^2p_{3/2}$,
$^1f_{5/2}$,  $^2p_{1/2}$, $^1g_{9/2}$) space. It is seen that the
predictions of DSM for $2\nu\;e^+$DBD half-lives are close to those of QRPA
and PHFB models. Extending the study  in \cite{y5} further, we have carried
out DSM calculations for $2\nu\;e^+$DBD half-lives for $^{74}$Se nucleus
and the results are reported in this brief report. We did not consider
$^{64}$Zn as  spherical shell model is well suited \cite{shzn64} for  the
three nuclei $^{64}$Zn, $^{64}$Cu and $^{64}$Ni due to the fact that they
are not well deformed with proton numbers close to the N=$28$ closed core.

Over the years, we have been using with success the deformed Shell Model
(DSM) based on Hartree-Fock states to study the spectroscopic properties,
such as band structures, shapes, nature of band crossings, electromagnetic
transition probabilities and so on, for medium heavy nuclei
\cite{sahu,npa96,br8082}.  More recently this model is applied to N=Z and
N=Z+1 nuclei by including isospin projection \cite{x3,x4}.  The
spectroscopic properties especially electromagnetic transitions like B(E2)
and B(M1) values provide a stringent test for the goodness of the nuclear
wave functions generated using the model.  It is also important to add that
DSM results are being used by many groups in the discussion of experimental
data for A $\sim$ 64-80 nuclei \cite{exp}. In addition DSM was used in
calculating transition matrix elements for $\mu-e$ conversion in $^{72}$Ge
\cite{app1} and in the analysis of data for inelastic scattering of 
electrons from $fp$-shell nuclei \cite{app3}. This model has also been used
for studying  $2\nu$ double beta decay transition matrix elements for
$^{76}$Ge $\rightarrow$ $^{76}$Se \cite{app2} with considerable success.
More recently in \cite{y5} we have applied DSM to study $\beta$-decay half
lives, GT distributions, electron capture rates and $2\nu\;e^+$DBD in
$^{78}$Kr.   All these confirm that DSM generates good nuclear wave
functions for nuclei in the mass region A $\sim$ 64-80.  Now, we will first
discuss the DSM formalism briefly and then the results for $^{74}$Se are
described.

\section{DSM formalism}

Half-life for the $2\nu\;e^+$DBD decay modes for the $0_I^{+} \rightarrow
J_F^{+}$ transitions, with $J_F^+$ for the daughter nucleus being $J_F^+ =
0^+_1$ or $2^+_1$, is given by
\begin{equation}
\left[ T_{1/2}^{2\nu }\left( k,J_F\right) \right] ^{-1}= G_{2\nu }
\left( k,J_F\right)\;
\left\vert M_{2\nu }(J_F)\right\vert ^{2}
\end{equation}
where $k$ denotes the modes $\beta ^{+}\beta ^{+}$, $\beta ^{+}$EC and
ECEC. As, besides the $0^+_1 \rightarrow 0^+_1$ transition, the ECEC mode
for $0^+_1 \rightarrow 2^+_1$ is also of experimental interest,  we are
considering both $J_F^+ = 0^+_1$ and $2^+_1$ in Eq. (1). The integrated
kinematical factors $G_{2\nu }\left( k,J_F\right)$ are independent of
nuclear structure  (except for the dependence on the excitation energy
$E_F$ of the $J_F$ state of the daughter nucleus) and they can be
calculated with good accuracy \cite{doi85, tom91, doi92, boe92}. Further,
the nuclear transition matrix elements (NTME) $M_{2\nu }$ are nuclear model
dependent and they are given by,
\begin{equation}
M_{2\nu }(J_F) = \displaystyle\frac{1}{\displaystyle\sqrt{J_F+1}}\;
\sum\limits_{N}\frac{\langle J_{F}^{+}||\mathbf{\sigma }
\tau^{-}||1_{N}^{+}\rangle \langle 1_{N}^{+}||\mathbf{\sigma }\tau
^{-}||0_{I}^{+}\rangle }{\left[E_{0}+E_{N}-E_{I}\right]^{J_F+1}}
\label{m2n}
\end{equation}
where $\left| 0_{I}^{+}\right\rangle ,\left| J_{F}^{+}\right\rangle $ and
$\left| 1_{N}^{+}\right\rangle $ are the initial, final and virtual
intermediate states respectively and $E_{N} (E_{I})$ is the energy of
intermediate (initial) nucleus. Note that $E_{0}=\frac{1}{2}\left(
E_{I}-E_{F}\right) =\frac{1}{2}W_{0}$. Here, $W_{0}$ is the total energy
released for different $2\nu\; e^+$DBD  modes. For $0^+_1 \rightarrow
0^+_1$ transitions  $W_{0}(\beta ^{+}\beta ^{+}) = Q_{\beta ^{+}\beta
^{+}}+2m_{e}$, $W_{0}(\beta ^{+}\mbox{EC}) = Q_{\beta ^{+}\mbox{EC}}+e_{b}$
and  $W_{0}(\mbox{ECEC}) = Q_{\mbox{ECEC}}-2m_{e}+e_{b1}+e_{b2}$. Note that 
the $Q$-values are given by the difference of neutral atomic masses of parent
and daughter nuclei involved in the positron double beta decay process and
$e_{b}$ is the binding energy of the absorbed atomic electron. For the
$0^+_1 \rightarrow 2^+_1$ ECEC transition, denoted by ECEC$^*$, we have
$W_{0}(\mbox{ECEC}^*) = Q_{\mbox{ECEC}}-\Delta{E} - 2m_{e} + e_{b1} +
e_{b2}$ where $\Delta{E}$ is the excitation energy of the $2^+_1$ state. We
have employed DSM to calculate the reduced matrix element appearing in Eq.
(\ref{m2n}).

In DSM, for a given nucleus, starting with a model space consisting of a
given set of single particle orbitals and effective two-body Hamiltonian,
the lowest prolate and oblate intrinsic states are obtained by solving the
Hartree-Fock (HF) single particle equation self-consistently. Excited
intrinsic configurations are obtained by making particle-hole excitations
over the lowest intrinsic state. These intrinsic states will not have good
angular momentum and good angular momentum states are obtained by angular
momentum projection from these intrinsic states. In general the projected
states with same $J$ but coming from different  intrinsic states will not
be orthogonal to each other. Hence  they are orthonormalized and then band
mixing calculations are performed. DSM is well established to be a
successful model for transitional  nuclei when sufficiently large number of
intrinsic states are included in the band mixing  calculations; see
\cite{y5} and references therein. Performing DSM calculations for the
parent, daughter and the intermediate odd-odd nucleus (here we need only
the $1^+$ states) and then using the DSM wavefunctions, the $\sigma\tau^-$
matrix elements in Eq. (2) are calculated. For further details see
\cite{y5}. Now we will discuss the results for $^{74}$Se.

\section{Results and discussion}

In our calculations of $^{74}$Se $2\nu\; e^+$DBD half-lives, for the
structure of the nuclei $^{74}$Se, $^{74}$As and $^{74}$Ge we have used a
modified  Kuo effective interaction \cite{int} in the ($^{2}p_{3/2}$,
$^{1}f_{5/2}$,  $^{2}p_{1/2}$, $^{1}g_{9/2}$) space with $^{56}$Ni as the
inert core. The single particle energies of these orbitals  are taken as
0.0, 0.78, 1.08 and 4.5 MeV respectively. DSM with modified Kuo effective
interaction has been quite successfully  used by us in describing many
important features of nuclei in A $\sim$ 60-80 region. In particular, shape
coexistence in spectra, observed $B(E2)$ values, band crossings and so on
in $^{70,72,74}$Se isotopes are well described by DSM \cite{y6}. Therefore,
just as $^{78}$Kr studied using DSM in \cite{y5}, for $^{74}$Se
$2\nu\;e^+$DBD decay DSM is expected to be good. We have also verified that
$^{74}$Ge spectroscopic properties are well described by DSM. For $2\nu\;
e^+$DBD half-lives calculations, we have first performed axially symmetric
HF calculations and obtained the lowest prolate HF intrinsic states.  The
lowest HF single particle  spectra for $^{74}$Se, $^{74}$Ge and $^{74}$As
nuclei are shown in Figs. 1, 2 and 3 respectively. Only prolate intrinsic
states are considered in these calculations and the oblate intrinsic states
are ignored just as in the previous $^{78}$Kr analysis \cite{y5} using DSM.
The reason for neglecting the oblate states has been discussed in an
earlier publication \cite{kct_rs}. For these three nuclei, we found that
the spectroscopic results obtained with only oblate states compare poorly
with experiment and hence we did not include oblate states in the final
calculation. We have also seen in the band mixing calculations that oblate
states do not mix with prolate states significantly and hence they are not
expected to affect our final results.  By particle-hole excitations from
the lowest intrinsic states shown in Figs. 1-3, excited  configurations are
generated. For  $^{74}$Se ground state $0^+$, 10 intrinsic states with
$K=0^{+}$ are used for band mixing. Similarly 24 configuration with
$K=0^{+}$ for $^{74}$Ge and  65 configurations with $K=1^{+}$ for $^{74}$As
are employed. We have verified that these configurations are sufficient to
provide adequate description of $2\nu\;e^+$DBD. Further increase in the
number of configurations does  not change the results significantly.

Using the wavefunctions generated by DSM, $2\nu $ $\beta ^{+}$EC and ECEC
half-lives for $^{74}$Se $\rightarrow $ $^{74}$Ge transitions are
calculated and the results are shown in Table I. The integrated kinematical
factors $ G_{2\nu }\left( k,J_F\right) $ have been calculated following the
prescription given by Doi and Kotani\cite{doi92}. The limits for $\beta
^{+}$EC processes in $^{74}$Se were determined only recently, in the
SuperNEMO project. Measurements of Se sample consisting of natural selenium
powder using a 400 cm$^3$ HPGe detector resulted in the first $T_{1/2}$
limits to be $>10^{18}-10^{19}$ yr \cite{se74} for $2\nu $ $\beta ^{+}$EC
and ECEC. The DSM results in Table I are the first theoretical  estimates
for the half-lives for positron double decay modes of $^{74}$Se and there
does not exist any other model calculations. Let us recall here the
statement in \cite{se74}: ``It is necessary to stress that $^{74}$Se has
never been investigated before and all results here are obtained for the
first time. Neither has this isotope been investigated theoretically; thus
there are no predictions with which to compare. Nevertheless, we will try
to estimate the significance of the obtained results and the possibility to
increase the sensitivity of this type of experiments in the future.''

\section{Conclusions}

In this brief report, by extending our recent results for $^{78}$Kr
\cite{y5}, we have presented results for positron double beta  decay half
lives for $^{74}$Se. They are obtained using the DSM model with a modified
Kuo interaction in ($^{2}p_{3/2}$, $^{1}f_{5/2}$,  $^{2}p_{1/2}$,
$^{1}g_{9/2}$) space. As spectroscopic properties of Se isotopes are well
described by DSM, the half-lives calculated for $2\nu\;e^+$DBD modes of
$^{74}$Se, given in Table I can be taken as reliable predictions. The
calculated half-life for the ECEC mode is $\sim 10^{26}$yr and it may be
possible to observe this in the near future with improved sensitivity of
experiments.

\acknowledgments

RS is thankful to DST (India) for financial support.

\pagebreak

\begin{sidewaystable}[tbp]

\caption{Experimental limit on half-lives T$_{2\nu }^{1/2}$ along with
theoretical estimates in Deformed Shell Model and corresponding phase space
factor G$_{2\nu }$ for possible decay modes for $^{74}$Se $\rightarrow 
^{74}$Ge. The Q- values are taken from 
\protect\cite{Aud03} and Abundance (P) values are from \protect\cite{Boh05}.
The range (a-b) given in parenthesis for the theoretical estimate of the
half-life is given for $g_A/g_V=1.261$ and $1$ respectively.}
\begin{tabular}{ccccccc}
\hline \hline \\
Transition & Q-value & P & Decay & G$_{2\nu }$ &
\multicolumn{2}{c}{$T_{2\nu}^{1/2}$ (in yrs)} \\ \cline{6-7}
\\
&  (in keV) & (in \%) & Mode & (in yr$^{-1}$) & Expt. limit & Theory
\\ \hline
\\
$^{74}$Se $\rightarrow ^{74}$Ge & 1209.7$\pm 0.6$ & 0.89 & $\beta ^{+}$EC 
& $\;\;\;\;2.05\times 10^{-29}$ & $\;\;\;\;>
1.9\times 10^{18}$\cite{se74} & $\;\;\;\;(14.99-37.9)\times 10^{30}$ \\ 
\\
&  &  & ECEC & $\;\;\;\;2.63\times 10^{-24}$ &  & $\;\;\;\;(7.56-19.12)
\times 10^{25}$\\
\\ 
&  &  & ECEC$^{*}$ & $\;\;\;\;3.06\times 10^{-27}$ & $\;\;\;\;>7.7
\times 10^{18}$\cite{se74} & $\;\;\;\;(15.55-39.32)\times 10^{30}$ \\ 
\\ \hline\hline\\
* represents g.s. to $2_{1}^{+}$ state transition ($\Delta{E} = 595.8$keV).
\end{tabular}
\end{sidewaystable}

\pagebreak

\newpage 

\begin{figure}[tbp]
\includegraphics[width=4.75in, height=7.5in]{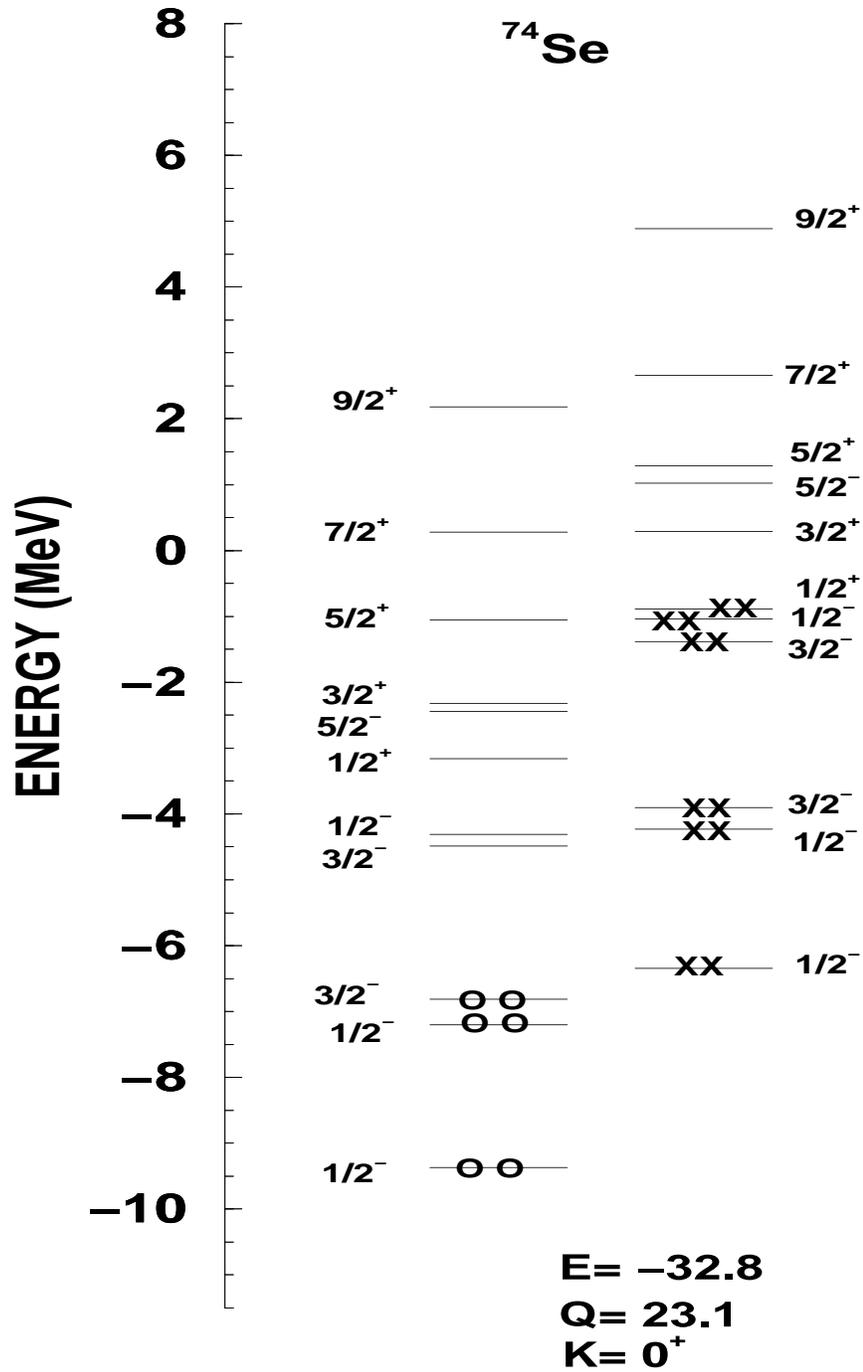}
\caption{HF single particle spectrum for $^{74}$Se. In the
figure circles represent protons and crosses represent neutrons. The 
Hartree-Fock energy ($E$) in MeV, mass quadrupole moment ($Q$) in units of 
the square of the oscillator length parameter and the total $K$ quantum number
of the lowest intrinsic state are given in the figure.}
\label{fig1a}
\end{figure}
\newpage 
\begin{figure}[tbp]
\includegraphics[width=5in, height=7.5in]{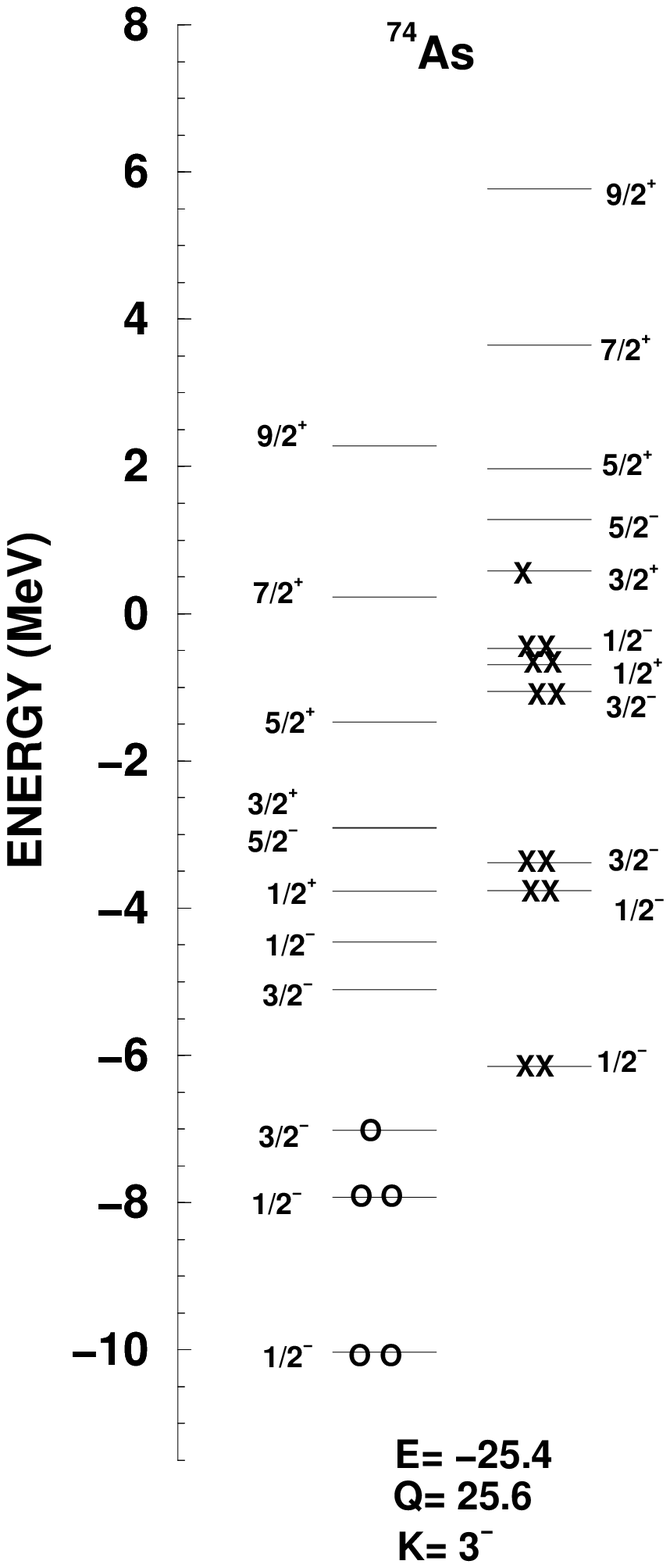}
\caption{Same as Fig. 1 but for $^{74}$As.}
\label{fig1b}
\end{figure}
\newpage 
\begin{figure}[tbp]
\includegraphics[width=5in, height=7.5in]{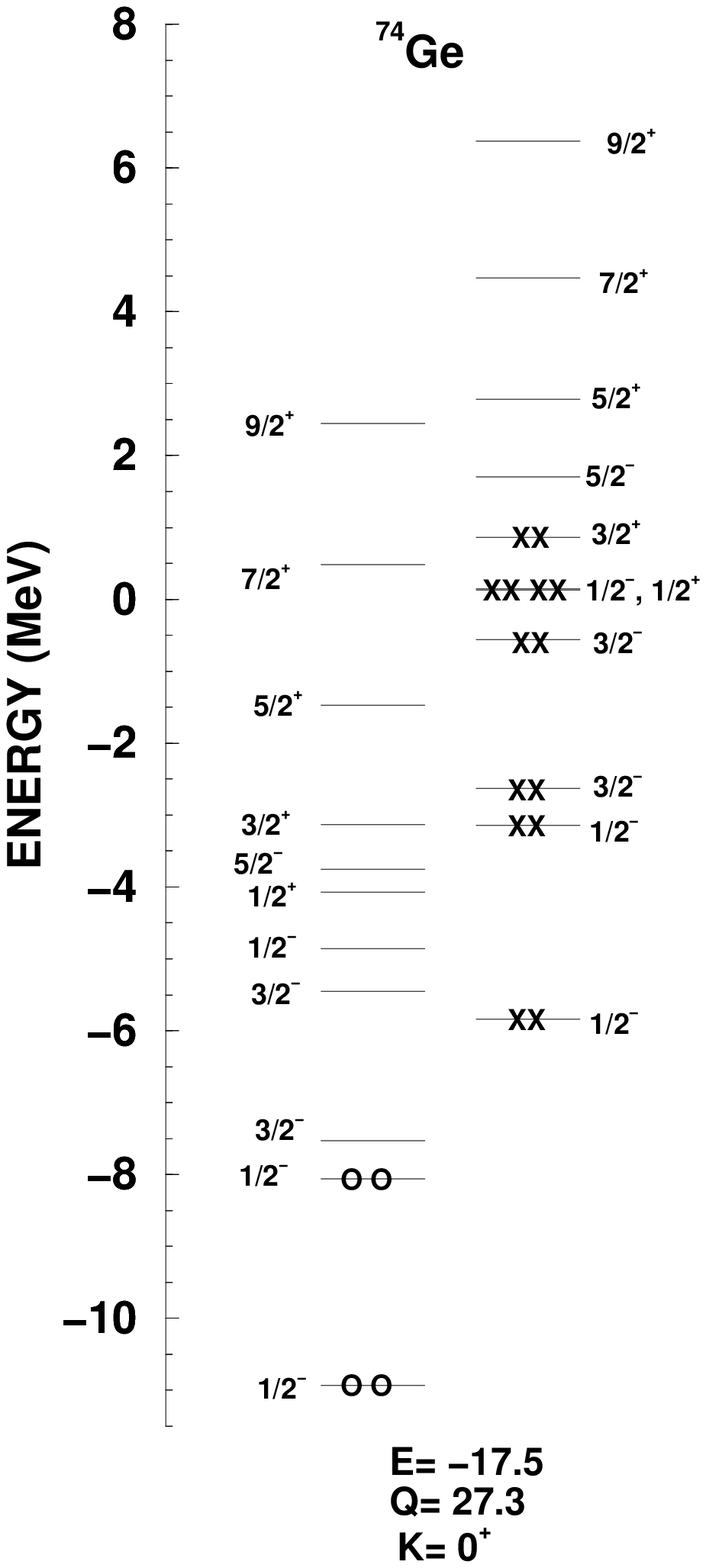}
\caption{Same as Fig. 1 but for $^{74}$Ge.}
\label{fig1c}
\end{figure}
\end{document}